\documentstyle[prb,aps]{revtex}
\begin{document}

\title{Why Self-Consistent Diagrammatic Perturbation Theory
 is ''just'' Perturbation Theory}
\author{Girish S. Setlur}  
\address{Department of Physics and Materials Research Laboratory,
 \\ University of Illinois at Urbana-Champaign , Urbana Il 61801 }
\maketitle

\begin{abstract}
 In this short write-up we argue that self-consistent diagrammatic perturbation
 theory(i.e. Feynman diagrams) for the one-particle Green function is 
 unable to capture some important qualitative features no matter how
 self-consistently the Green functions are obtained. This write-up is intended
 to highlight the short-comings of perturbation theory and also tout the
 advantages of the sea-boson technique(hep-th/9706006). 
\end{abstract}

\section{The Arguments}

 In this write-up, the advantages of the sea-boson technique\cite{Setlur}
 are highlighted and contrasted with the short-comings of
 self-consistent diagrammatic perturbation theory(i.e. Feynman diagrams
 or its algebraic counterpart introduced by Schwinger\cite{Baym}).
 For more details the reader is refered to our published work
 \cite{Setlur}. Here and henceforth the term ''the text'' refers to the
 book by Kadanoff and Baym\cite{Baym}. For the sake of definiteness we shall 
 focus here only on fermions and that too without spin(S=0)
 therefore we shall employ only one of the
 signs used in the text namely the lower sign corresponding to fermions. 
 All equation numbers starting with the letters ''KB'' are from this book. 
 Let us start with the equation of motion for the Green function(KB 12-7). 
\begin{equation}
(i\frac{ \partial }{ \partial t_{1} }
 + \frac{ \nabla^{2}_{1} }{2m} - U_{eff}(1))
 G(1,1^{'};U) = \delta(1-1^{'})
 + \int d2\mbox{       }\Sigma^{'}(1,2;U_{eff})G(2,1^{'};U_{eff})
\label{GREENEQN} 
\end{equation}	
 where,
\begin{equation}
 U_{eff}(1) = U(1) + \int d2\mbox{   } v(1-2)[-i\mbox{  }G(2,2^{+};U_{eff})
-n_{0}] 
\end{equation}
here $ n_{0} $ is the mean density of particles.
\begin{equation} 
\Sigma^{'}(1,1^{'};U_{eff}) = i\mbox{  }v_{S}(1,1^{'};U_{eff})
G(1,1^{'};U_{eff})
 + i\int d3\mbox{   }d4\mbox{   }
v_{S}(1,3;U_{eff})G(1,4;U_{eff})
\frac{ \delta \Sigma^{'}(4,1^{'};U_{eff}) }
{\delta U_{eff}(3) }
\label{SELFEN}
\end{equation}
 The $ v(1-2) = \lambda\mbox{     }v({\bf{x}}_{1}-{\bf{x}}_{2})\delta(t_{1}-t_{2}) $
 is the bare Coulomb interaction and $ G(1,2;U=0) $ is the
 full one-particle Green function in equilibrium and
 $ \lambda $ is a dimensionless coupling constant. 
 Further(KB 12-10),
\[
v_{S}(1,3) = v(1-3) -i\int d2\mbox{       }d4\mbox{    }
v_{S}(1,2) G(4,2)G(2,4^{+})v(4-3)
\]
\begin{equation}
-i \int v_{S}(1,2)G(4,5)\frac{ \delta \Sigma^{'}(5,5^{'}) }
{\delta U_{eff}(2) }
G(5^{'},4) v(4-3)
\end{equation}
 As it stands the above sets of equations are defined independent of
 perturbation theory. That is, $ \lambda $ is not assumed to be small.
 We wish to argue that within the framework of perturbation theory,
 the Green function obtained from the above equations has a simple analytic
 structure in the vicinity of the origin of the coupling constant space
 ($\lambda = 0$ ).
 More precisely, we argue that, following the letter and spirit of 
 perturbation theory leads us to the inevitable conclusion that the
 one-particle Green function is an analytic function of the coupling
 constant, perhaps with zero radius of convergence. This is true even if
 the Green function is obtained ''self-consistently'' (that is, by solving
 a system of coupled non-linear integro-differential equations).
 This state of affairs should be contrasted with the sea-boson method where
 we found that at least in one-dimension, the one-paricle Green function
 has a nonanalytic dependence in the coupling at the origin of the
 coupling constant space\cite{Setlur}.
 Thus this feature is completely erased 
 from diagrammatic perturbation theory due to the latter's inherent
 and as it happens, drastic assumptions. 
 The letter and spirit of perturbation theory demands that we solve for
 the correlation self-energy by expanding in powers of the (screened
 or shielded) Coulomb interaction. When this is done, it may be seen quite
 easily that the correlation self-energy is an
 analytic function of the coupling constant
 at the origin of the the coupling constant space. To see this more clearly
 let us first observe that the solutions to Eqs.(~\ref{GREENEQN})
 and (~\ref{SELFEN}) together with KMS boundary conditions \cite{Baym}
 are unique. Therefore the Green functions and self-energies 
 possess unique analytic structures. 
 Furthermore, if an analysis shows that a Green function is (and
 self-energy) analytic at the origin of the coupling
 constant space is consistent with
 the above sets of equations then indeed this is the only possible analytic
 structure possessed by these quantities. This is what we shall now 
 argue. Let us retain only the GW-part of the correlation self-energy,
 that is, neglect all derivatives of the self-energy with respect to the
 effective potential as suggested by Kadanoff and Baym\cite{Baym}.
 When this is done
 one may proceed to convince oneself of the analyticity of the Green function
 as follows. Since we are interested in the neighbourhood of $ \lambda = 0 $,
 one can argue that the zeroth order approximation to the full Green function
 is $ G_{0}(1,1^{'}) $, namely the noninteracting one. Then one may use
 this to evaluate the self-energy and the screened Coulomb interaction.
 Note that so long as we are sufficiently close to $ \lambda =0 $, 
 no loss of generality is entailed by this procedure. In particular,
 a ''self-consistent'' solution to this system for $ \lambda $ sufficently 
 close to $ \lambda = 0 $ is equivalent to the usual pertubative procedure
 we just indicated. To proceed with the description of
 the iterative procedure, then one may re-evaluate the full Green function
 by including linear terms in $ \lambda $ on the right-hand side of
 Eq.(~\ref{GREENEQN}). This procedure when repeated leads us to the
 unambiguous conclusion that the full Green function is indeed an analytic
 function of $ \lambda $ at $ \lambda = 0 $. Two points are worth stressing
 again. First is that this conclusion depends rather strongly on the 
 fact that we have chosen to ignore the derivatives of the self-energy
 with respect to the effective potential. Only then are we able to reduce
 the system  to a familiar set of integro-differential equations rather
 then a set of functional equations. The second is that having convinced
 ourselves that a Green function that is analytic at $ \lambda = 0 $
 is consistent with the above sets of equations, we are also led
 to the conclusion that this structure is unique since
 the Green function itself is unique. Thus if a skeptical reader tries to argue
 that maybe if one had started with an ansatz for the full Green function
 that is non-analytic in $ \lambda $ then one would have found that this is
 also consistent with the above sets of equations. Not so. The reason as
 we just alluded to is uniqueness. Having said this, we would now like to 
 highlight some of the advantages of the sea-boson method that we
 have introduced into the literature\cite{Setlur}. There we found that
 for small enough values of the coupling, the momentum distribution(or the 
 equal-time component of the full Green function) is nonanalytic
 in the coupling, thus demonstrating that the sea-boson method is superior.
 For the sake of completeness, we reproduce the results here. 
\[
\langle c^{\dagger}_{ k }c_{ k } \rangle
 = n_{F}(k) +
(2\pi k_{F})
\int_{-\infty}^{+\infty} \mbox{ }\frac{ dq_{1} }{2\pi}\mbox{ }
\frac{ \Lambda_{ k - q_{1}/2 }(-q_{1}) }
{ 2\omega_{R}(q_{1})(\omega_{R}(q_{1}) + \omega_{k - q_{1}/2}(q_{1}))^{2}
(\frac{ m^{3} }{q_{1}^{4}})( cosh(\lambda(q_{1})) - 1 ) }
\]
\begin{equation}
- (2\pi k_{F})
\int_{-\infty}^{+\infty} \mbox{ }\frac{ dq_{1} }{2\pi}\mbox{ }
\frac{ \Lambda_{ k + q_{1}/2 }(-q_{1}) }
{ 2\omega_{R}(q_{1})(\omega_{R}(q_{1}) + \omega_{k + q_{1}/2}(q_{1}))^{2}
(\frac{ m^{3} }{q_{1}^{4}})( cosh(\lambda(q_{1})) - 1 ) }
\end{equation}
here,
\begin{equation}
\lambda(q) = (\frac{2 \pi q}{m})(\frac{1}{v_{q}})
\end{equation}
\begin{equation}
\omega_{R}(q) = (\frac{ |q| }{m})
\sqrt{ \frac{ (k_{F} + q/2)^{2}  - (k_{F} - q/2)^{2}exp(-\lambda(q)) }
{ 1 - exp(-\lambda(q)) } }
\end{equation}
\begin{equation}
n_{F}(k) = \theta(k_{F} - |k|), \mbox{            }
\Lambda_{ k }(q) = n_{F}(k + q/2)( 1 - n_{F}(k - q/2))
\end{equation}
 The interaction in the denominator present in the quantity $ \lambda(q) $ 
 means that the momentum distribution is nonanalytic in the coupling
 at $ v_{q} = 0 $, thus making our claims concrete. 
 
 The author wishes to thank Mr. Erich Mueller for critical comments 
 that led to this write-up. Thanks are also due to Prof. J.J. Song and
 her group at Oaklahoma State University for their support both financially
 and morally. The author may be contacted at setlur@mrlxpa.mrl.uiuc.edu.

\end{document}